\documentclass[%
 pra,reprint,
 showpacs,
 amsmath,amssymb,
 aps,
]{revtex4-1}

\usepackage{graphicx}
\usepackage{dcolumn}
\usepackage{bm}
\usepackage{color}
\usepackage{textpos}
\usepackage{dsfont}

\renewcommand{\_}[1]{_\mathrm{ #1 }}
\newcommand{\m}[1]{\mathrm{ #1 }}

\newcommand{\ket}[1]{| #1 \rangle}
\newcommand{\bra}[1]{\langle #1 |}

\renewcommand{\i}[0]{\mathrm{i}}
\newcommand{\e}[0]{\mathrm{e}}
\renewcommand{\[}[0]{\left[}
\renewcommand{\]}[0]{\right]}
\renewcommand{\(}[0]{\left(}
\renewcommand{\)}[0]{\right)}

\newcommand{\expv}[1]{\langle #1 \rangle}

\begin{document}

\preprint{APS/123-QED}

\title{Master equation for high-precision spectroscopy }

\author{Andreas Alexander Buchheit}
\email{andreas.buchheit@gmail.com}
\author{Giovanna Morigi}%
 \affiliation{%
Theoretische Physik, Universit\"at des Saarlandes, D-66123 Saarbr\"ucken, Germany
}%

\date{\today}%

\begin{abstract}
The progress in high-precision spectroscopy requires one to verify the accuracy of theoretical models such as the master equation describing spontaneous emission of atoms. For this purpose, we apply the coarse-graining method to derive a master equation of an atom interacting with the modes of the electromagnetic field. This master equation naturally includes terms due to quantum interference in the decay channels and fulfills the requirements of the Lindblad theorem without the need of phenomenological assumptions. We then consider the spectroscopy of the 2S-4P line of atomic Hydrogen and show that these interference terms, typically neglected, significantly contribute to the photon count signal. These results can be important in understanding spectroscopic measurements performed in recent experiments for testing the validity of quantum electrodynamics.
\end{abstract}


\maketitle

\section{Introduction}
For many decades the spectroscopic properties of laser-driven atoms have been successfully modeled by a Born-Markov master equation for the electronic degrees of freedom, where spontaneous decay is described by a Liouvillian term and the frequency shifts induced by the coupling with the field are included as corrections to the Hamiltonian \cite{Landau,Milonni,breuer2010theory,agarwal2012quantum}. The resulting Optical Bloch Equations (OBE), namely, the equations of motion for the density-matrix elements, are by now a tool which is routinely employed to interpret the experimental curves \cite{Allen1975}. 

Recent spectroscopic measurements, pushing the precision to the limits, reported discrepancies with the predictions of the OBE, which, if verified, could have consequences on the validity of quantum electrodynamics \cite{pohl2010size,precisioncritique}.  It has been conjectured that  these discrepancies could emerge from contributions to the OBE, which have been identified in the original derivations starting from a fully quantized description of the modes of the electromagnetic field \cite{Milonni} but have been neglected so far \cite{Horbatsch2010}. These terms describe interference between decay channels and are usually referred to as cross-damping terms \cite{Cardimona1983,ficek2005}. They are important in order to correctly reproduce radiation damping in a harmonic oscillator \cite{Cook1984} but are typically discarded in the OBE used for atomic spectroscopy, since they were assumed to be negligible. For anharmonic spectra, moreover, their inclusion in the master equation used in atomic spectroscopy \cite{Milonni,agarwal2012quantum,Cohen-Tannoudji1998,Carmichael} leads to a form which does not fulfil the requirements of the Lindblad theorem \cite{breuer2010theory}, unless additional assumptions are made \cite{ficek2005}. For closed level structures it was argued that these terms could give rise to "steady-state quantum beats" \cite{Cardimona1982steady} and in general to interference effects in the radiative emission of parallel quantum dipoles \cite{Swain,Scully1996,Keitel,Berman1998}. Studies which analysed the effects of the cross-damping terms for spectroscopy used phenomenological models \cite{Horbatsch2010} or derived scattering amplitudes \cite{jentschura2002nonresonant,brown2013quantum,YostInterference2014}, arguing that cross-damping could be responsible for systematic shifts of the order of kHz in the spectroscopy of atomic Hydrogen. This situation motivates a systematic theoretical derivation, which delivers a valid master equation and which can thus allow one to quantitatively predict whether and how such interference terms affect the spectroscopic measurements. 

In this work we  derive a master equation for the electronic bound states of an atom coupled to the quantum electromagnetic field by applying the coarse-graining procedure developed in Refs. \cite{lidar1999,lidar2001,majenz2013}, with some appropriate modifications.
Differing from previous treatments \cite{Milonni,ficek2005}, the coarse-graining approach allows us to consistently include terms due to the interference in the radiative processes of atomic transitions: These terms preserve the Lindblad form and their coefficients do not depend on the particular choice of the coarse-graining time step, provided this is chosen within the range of validity of the Born-Markov approximation. However, the coarse-graining procedure as in Ref.~\cite{majenz2013} delivers an involved form of the master equation, where the physical origin of the individual terms is non evident and where the coefficients are transcendental functions of the atomic parameters. Here, by an appropriate modification of the derivation we find a simpler form which can be set in connection to and compared with master equations applied so far for atomic spectroscopy. This allows us to quantify the contribution of the cross-damping interference in the spectroscopic measurements, as we show below for the specific case of the 2S-4P transition in atomic Hydrogen, a case in point for the proton size puzzle \cite{pohl2010size}. 

This article is organized as follows. In Sec.~\ref{Sec:2} we introduce the Hamiltonian and derive a master equation in Lindblad form using a modified coarse-graining procedure. We discuss the resulting cross-damping and cross-shift terms and set them into connection with previous results in the literature. In Sec.~\ref{Sec:3} we provide a concrete case study by applying the master equation we derive to 2S-4P spectrosopy in atomic Hydrogen, while in Sec. \ref{Sec:Time} we calibrate the coarse-graining time. The conclusions are drawn in Sec. \ref{Sec:Conclusions}, while the Appendix complements the discussion in Sec. \ref{Sec:3}.

\section{Derivation of the master equation}
\label{Sec:2}

In order to clarify the origin of the cross interference terms and the problem with their systematic treatment in the literature, we report the derivation of the coarse-grained master equation as in Refs. \cite{lidar1999,lidar2001,majenz2013}. We aim at deriving the master equation for the reduced density matrix $\hat \rho(t)$ at time $t$ of a valence electron of mass $m$ bound to an atom and coupled to the modes of the electromagnetic field (EMF). 

In this treatment, we assume the electronic energies to be discrete and thus neglect the continuum spectrum of ionization. We furthermore restrict ourselves to a finite number of $N$ levels assuming that the occupation of the remaining states is negligible at all times. This is reasonable for sufficiently weak exciting fields \cite{Cohen-Tannoudji1998}. Moreover we neglect the center-of-mass motion of the atom and associated effects like the Doppler shift \cite{Allen1975}.

\subsection{System model and Hamiltonian}

We first consider the density matrix $\hat\chi(t)$ of the composite atom and EMF system, from which the operator $\hat \rho(t)$  is obtained after tracing $\hat\chi(t)$ over the EMF degrees of freedom, $\hat{\rho}(t)={\rm Tr}\_R\{ \hat \chi(t)\}$. The density matrix $\hat \chi(t)$ undergoes a coherent dynamics determined by the Hamiltonian 
\begin{equation}
\Hat{H}=\Hat{H}\_A  + \Hat{H}_{\m R}+\Hat V\,,
\end{equation}
according to the von-Neumann equation 
\begin{equation}\partial_t \hat \chi=[\hat H,\hat\chi]/{\i\hbar}\,,
\end{equation}
with $\hbar$ the reduced Planck constant. Here, $\Hat{H}_A$ is the atomic part of the Hamiltonian and satisfies the eigenvalue equation $$\Hat{H}\_A \ket{n}:=E_n \ket{n}\,,$$ with $\ket{n}$ the eigenstates of $\Hat H\_A$, representing the bound states of the valence electron, and $E_n$ their associated energies. The free evolution of the quantized electromagnetic field (EMF) is described by Hamiltonian $$\Hat{H}_{\m R}=\sum_{\lambda} \hbar \omega_\lambda \Hat{a}^{\dagger}_{\lambda} \Hat{a}_{\lambda}\,,$$ with the annihilation and creation operators $\Hat{a}_\lambda$ and $\Hat{a}^{\dagger}_\lambda$ of a photon in the field mode at frequency $\omega_\lambda$, wave vector $\vec{k}$ and transverse polarization $\vec{e}_\mu (\vec{k}) \perp \vec{k}_\lambda$ with $\mu \in \{-1,1\}$ ($\lambda:=(\vec{k},\mu)$, $\omega_\lambda=c|\vec{k}|$) \cite{Cohen-Tannoudji1998} (we drop the energy of the vacuum state). Here, $[\hat a_\lambda, \hat a_{\lambda'}^\dagger]=\delta_{\lambda,\lambda'}$. Moreover, the sum in $\Hat{H}_{\m R}$ is restricted to modes with $\omega_\lambda<\omega\_{cut}$, with  $\omega\_{cut}\sim mc^2/\hbar$ the cutoff frequency, $m$ the electronic mass, and $c$ the speed of light.

The interaction between the electromagnetic field and the electronic transitions of the atom in the long-wavelength approximation is given in the dipole representation and can be cast in the form
\begin{equation}
\hat V =\hbar \sum_i (\hat \Gamma_i^\dagger \hat \sigma_i+\hat \Gamma_i \hat \sigma_i^\dagger)\,,
\label{eq:interactionhamiltonian}
\end{equation}
where  $\hat \sigma_i=\ket{i_1}\bra{i_2}$ projects a state $i_2$ to a lower lying state $i_1$, coupled by a dipolar transition with moment $\vec{d_i}=\langle i_1|\hat{\vec{d}}|i_2\rangle$ and frequency $\omega_i=(E_{i_2}-E_{i_1})/\hbar$. Moreover, 
\begin{equation}
\hat \Gamma_i = \sum_\lambda \( g_i^\lambda \hat a_\lambda+{\bar g_i^{\lambda^*}} \hat a_\lambda^\dagger \)\,,
\end{equation} 
where  $g_{i}^\lambda:=-\m{i} \sqrt{\frac{2 \pi \omega_\lambda}{\hbar \mathcal V }}\vec{d}_{i}^*\cdot \vec{\m{e}}_\lambda$ and $\bar g_{i}^\lambda:=-\m{i} \sqrt{\frac{2 \pi \omega_\lambda}{\hbar  \mathcal V }}\vec{d}_{i}\cdot \vec{\m{e}}_\lambda$ are the coupling strength and $ \mathcal V$ is the quantization volume.
The long-wavelength approximation limits the validity of Eq. \eqref{eq:interactionhamiltonian} to low-lying atomic states, where the size of the bound electron wave packet is smaller than the optical wavelength. \vspace{0.2cm}

\subsection{Derivation of the Liouville Equation}

We now cast the dynamics in the interaction picture with respect to the unperturbed Hamiltonian $\Hat{H}_0=\Hat{H}\_A +\hat H\_R$. In particular, given $\hat{X}(t)$ an operator in the laboratory frame, we denote by $\tilde{X}(t)$ the corresponding operator in interaction picture, such that
\begin{equation}
\tilde{X}(t)=\hat U_I(t)\hat X(t)\hat U_I^\dagger(t)\,,
\end{equation}
with $\hat U_I(t)=\exp\left(\i \hat H_0t/\hbar\right)$.
In the interaction picture the formal time evolution of the density matrix $\tilde \chi(t')$ for times $t'>t$ reads $$\tilde \chi(t')=\tilde U(t,t') \tilde \chi(t) \tilde U^\dagger (t,t')\,,$$ 
where $$\tilde U(t,t')={\mathcal T}\exp\left(-\i \int_t^{t'}{\rm d}\tau\tilde{V}(\tau)\right)$$ is the time evolution operator in interaction picture and ${\mathcal T}$ denotes the time ordering. 

For $t'=t+\Delta t$, with $\Delta t>0$ and sufficiently small, we expand the right-hand side to second order in the interaction $\tilde V$, assuming that the coupling between atom and field is weak. We take the partial trace over the EMF-modes of the resulting equation and get $$\tilde \rho(t+\Delta t)=\tilde \rho(t)+\Delta t\mathcal{L} \tilde \rho(t)\,,$$ where
\begin{equation}
\label{M:Eq}
\mathcal{L}  \tilde \rho(t)=\frac{1}{{\rm i}\hbar}\left[ \tilde{V}_R(t),\tilde{\rho}(t)\right]+\frac{1}{{\rm i}\hbar}\left[\tilde H_\m{S}(t),\tilde \rho(t) \right]
+\mathcal{L}\_D(t)\tilde \rho(t)\,,
\end{equation}
with $\tilde V_R(t)$ and $\tilde H_\m{S}(t)$ Hermitian operators, and $\mathcal{L}\_D(t)$ a superoperator of Lindblad form \cite{breuer2010theory}, whose detailed forms are given in what follows.

\begin{figure}
	\begin{center}
		\includegraphics[width=0.4\textwidth]{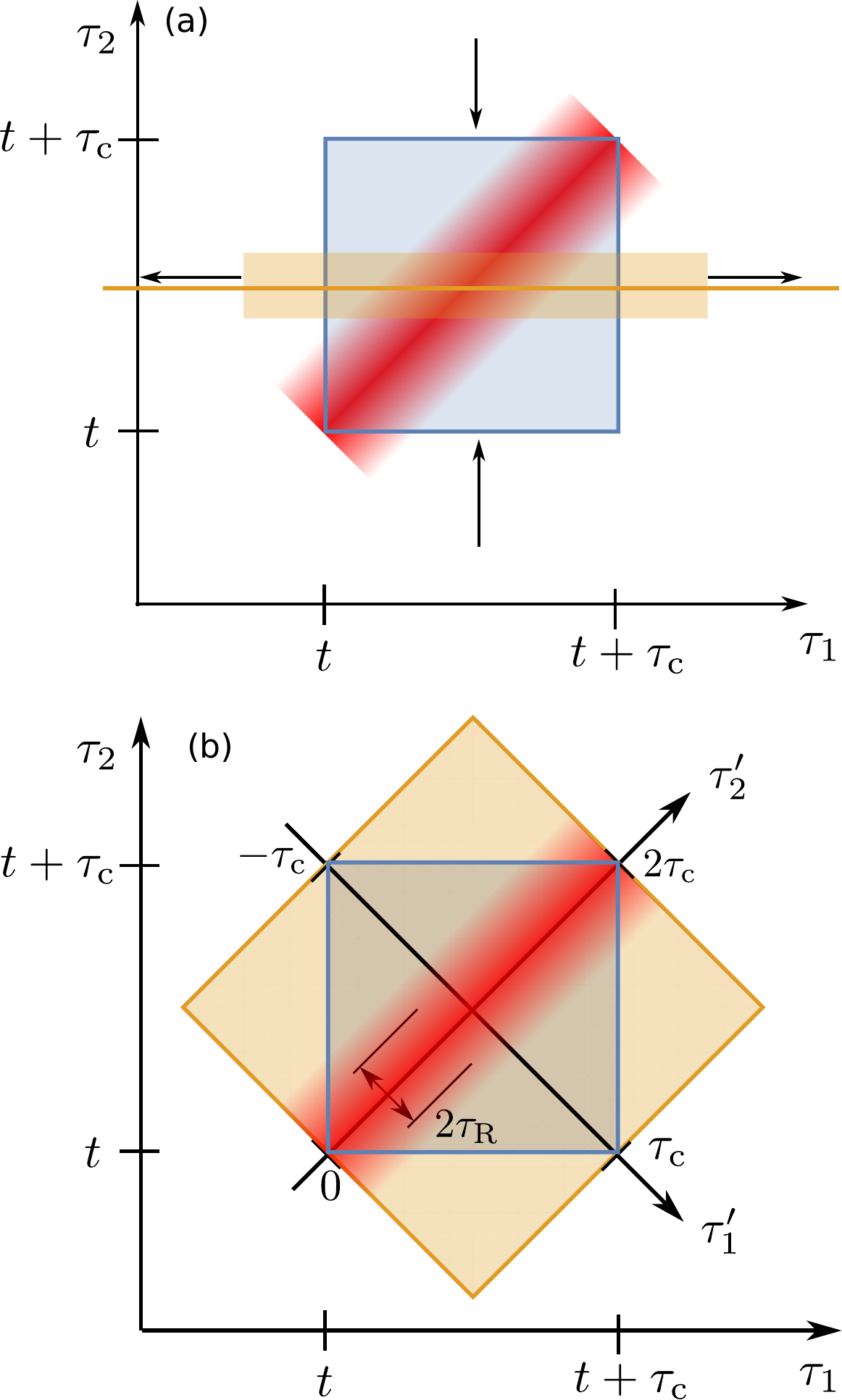}
	\end{center}
	\caption{Time integration regions used to evaluate Eqs. \eqref{H}-\eqref{L}. The blue rectangle is the region with both $\tau_1$ and $\tau_2$ integrated over $[t,t+\tau\_c]$.  (a) Integration region leading to Eq. \eqref{eq:dissipatorficek} after extending the integration over $\tau_1$ to $(-\infty,\infty)$, corresponding to the limit $\tau\_c/\tau\_R \to \infty$, while shrinking the integration interval over $\tau_2$, corresponding to $\tau\_c/\tau\_A \to 0$ (see arrows). (b) New integration region: The yellow box indicates the region with  $\tau_1'=\tau_1-\tau_2\in [-\tau\_c,\tau\_c]$ and $\tau_2'=\tau_1+\tau_2-2t\in [0,2\tau\_c]$, which preserves the symmetries of the cross-damping coefficients, thus warranting the Lindblad form. The blue and yellow integration regions are equivalent, being the correlation function  in the integrand different from zero only within the red small stripe of width $\sim 2\tau\_R\ll \tau_c$. }
	\label{Fig:1}
\end{figure}

So far, Eq. \eqref{M:Eq}  provides a valid description of the dynamics only in the interval $[t,t+\Delta t]$ and only when the time step $\Delta t$ is sufficiently short so to warrant the validity of perturbation theory. Moreover, we have asssumed that at the initial time $t$ the density matrix is $\tilde{\chi}(t)=\tilde{\rho}(t)\otimes \mathcal R(t)$, with $\mathcal R(t)$ the reservoir density matrix. Specifically, 
\begin{equation}
\tilde V_R(t)={\rm Tr}_R\left\{\mathcal R(t)\int_t^{t+\Delta t}{\rm d}\tau \tilde{V}(\tau)/\Delta t\right\}
\end{equation}
and
\begin{widetext}
\begin{eqnarray}
\label{H}
&&\tilde H_\m{S}(t)=-\frac{\rm i}{2\hbar\Delta t}\int_t^{t+\Delta t}{\rm d}\tau_1\int_t^{t+\Delta t}{\rm d}\tau_2 \theta(\tau_1-\tau_2){\rm Tr}_R\left\{ \left[\tilde{V}(\tau_1),\tilde{V}(\tau_2)\right]\mathcal R(t)\right\}\,,\\
&&\mathcal{L}_D(t)\tilde \rho(t)=\frac{1}{2\hbar^2\Delta t}\int_t^{t+\Delta t}{\rm d}\tau_1\int_t^{t+\Delta t}{\rm d}\tau_2{\rm Tr}_R\left\{ 2\tilde{V}(\tau_1)(\tilde\rho(t)\otimes \mathcal R(t)) \tilde{V}(\tau_2)-[\tilde{V}(\tau_1)\tilde{V}(\tau_2),\tilde\rho(t)\otimes \mathcal R(t)]_+\right\}\label{L}\,,
\end{eqnarray}
with $\theta(t)$ Heaviside's function and $[,]_+$ the anticommutator. 

We now assume that the bath is at equilibrium and its correlation time $\tau_R$ is orders of magnitude smaller than the typical time scale $\tau_A$ of the atom, which allows us to perform the Markov approximation. In this specific limit Eq. \eqref{M:Eq} can be cast in the form of a differential equation, which is valid at all times $t'>t$, provided that the time step $\Delta t$ determining the coarse-graining of the time evolution can be chosen to be $\Delta t=\tau_c$ with $\tau_R\ll\tau_c\ll\tau_A$ \cite{Carmichael,lidar2001}. For this system this inequality is fulfilled since it is reasonable to assume that the EMF is in a thermal state $\mathcal R(t)=\mathcal R_0=\exp(-H_R/k_BT)/Z$ with temperature $T\sim 300$K and the partition function $Z={\rm Tr}\{\exp(-H_R/k_BT)\}$, giving $\tau\_R\simeq\hbar/(\m{k}\_B T)\sim 10^{-13}$sec \cite{Carmichael}. Then,
\begin{equation}
\partial_{t'}\tilde \rho(t')\equiv(\tilde \rho(t'+\tau\_c)-\tilde \rho(t'))/\tau_c=\mathcal{L}(t')\tilde \rho(t')\,,
\end{equation}
with $\mathcal{L}(t)$ in Eq.~\eqref{M:Eq}.  This equation is valid for any coarse-grained time $t'>t$~\cite{Cohen-Tannoudji1998,schaller2008preservation}. 

We now determine the explicit form of the dissipator using Eq. \eqref{eq:interactionhamiltonian} in Eq. \eqref{L}:
\begin{eqnarray}
\mathcal{L}_{D}(t) \tilde \rho(t)&=&\frac{1}{2 \tau\_c}\sum_{i,j}\int_t^{t+\tau\_c}\m d \tau_1 \, \int_t^{t+\tau\_c} \m d \tau_2\, \Bigg[  C(\tau_1-\tau_2)\bigg\{\vec d_i \cdot \vec d_j[\tilde  \sigma_j(\tau_2)\tilde  \rho(t),\tilde\sigma_i(\tau_1)]+\vec d_i^*\cdot \vec d_j [\tilde \sigma_j(\tau_2)\tilde  \rho(t),\tilde \sigma_i^\dagger(\tau_1)]\bigg\}\notag \\ 
&&+C(\tau_1-\tau_2)\bigg\{\vec d_i^* \cdot d_j[\tilde \sigma_j^\dagger(\tau_2) \tilde  \rho(t), \tilde \sigma_i(\tau_1)]+\vec d_i\cdot \vec d_j[\tilde \sigma_j^\dagger(\tau_2) \tilde  \rho(t),\tilde \sigma_i^\dagger(\tau_1)] \bigg\} + \m{H.c.}\Bigg]\,.
\label{eq}
\end{eqnarray}
Here we have introduced the correlation function 
\begin{equation}
C(\tau)=2/(3\pi\hbar c^3) \int_0^{\omega\_{cut}} \m d \omega \, \omega^3 \Big( \[1+n(\omega,T)\]\e^{-\i\omega \tau}+n(\omega,T) \e^{\i \omega \tau}\Big)\,,
\end{equation}  
where $n(\omega,T)$ is the mean photon number at frequency $\omega$, and is negligible for optical frequencies and room temperature. 
We note that the products $\vec d_i^* \cdot \vec d_j$ and $\vec d_i \cdot \vec d_j$ are real, since the elements of each dipole moment are real in the spherical basis. The terms proportional to $\vec d_i^*\cdot \vec d_j$ give the main contribution whereas the nonsecular terms proportional to $\vec d_i \cdot \vec d_j$ are oscillating at sums of transition frequencies in the interaction picture and  lead to corrections that are of the same order of magnitude as the error of the perturbative expansion itself \cite{nonrotatingwave,AgarwalNonSec}. We thus drop the nonsecular terms containing $\hat \sigma_i\hat \sigma_j$ and retain the secular terms, casting the dissipator in the form 
\begin{eqnarray}
\mathcal{L}_D(t)\tilde \rho(t)&=&\frac{1}{2\tau\_c}\sum_{i,j}\int_t^{t+\tau\_c}\m d \tau_1 \, \int_t^{t+\tau\_c} \m d \tau_2\, \vec d_i^*\cdot \vec d_j 
\Bigg[C(\tau_1-\tau_2) \e^{\i \omega_i (\tau_1-t)} \e^{-\i\omega_j( \tau_2-t)}\Big\{\[\tilde \sigma_j(t) \tilde \rho(t), \tilde \sigma_i^\dagger(t)\]+\[\tilde \sigma_j(t) , \tilde \rho(t) \tilde \sigma_i^\dagger(t)\]\Big\} \notag \\
&&+C(\tau_1-\tau_2) \e^{-\i \omega_i (\tau_1-t)} \e^{\i\omega_j (\tau_2-t)}\Big\{\[\tilde \sigma_j^\dagger(t) \tilde \rho(t), \tilde \sigma_i(t)\]+\[\tilde \sigma_j^\dagger(t) , \tilde \rho(t) \tilde \sigma_i(t)\]\Big\}\Bigg]\,,
\label{eq:dissipator112}
\end{eqnarray}
where we used that $\tilde \sigma_j^\dagger(\tau_1)= \e^{\i\omega_j (\tau_1-t)}\tilde \sigma_j^\dagger(t)$. The final master equation is then found by evaluating the coefficients, thus performing the integral. In the present form,
however, the coefficients in Eq.~\eqref{eq:dissipator112} are transcendental functions of the atomic parameters, and cannot be simply compared with the master equations used so far in the literature \cite{Milonni,ficek2005}. 

Before proceeding, it is now useful to discuss the master equation for spectroscopy as in Refs. \cite{Milonni,Cardimona1983,ficek2005}, when one includes terms scaling with $\hat \sigma_i^{\dagger} \hat \sigma_j$ (for $i\neq j$), which we denote by cross interference terms. Its structure is similar to the one of Eq.~\eqref{M:Eq}, however the dissipator reads \cite{Milonni,Cardimona1983,ficek2005}:
\begin{align} 
\mathcal{L}_{D}^F \hat \rho(t) = \hspace{-0.5mm} \sum_{i,j}
(1+n(\omega_{j},T)) \bigg(\frac{\gamma_{ij}^\m{F}}{2}\[\hat \sigma_j \hat \rho(t), \hat \sigma_i^\dagger\]+\frac{\gamma_{ji}^\m{F*}}{2}\[\hat \sigma_j , \hat \rho(t) \hat \sigma_i^\dagger\]\bigg) \hspace{-1 mm}+\hspace{-0.5mm}\sum_{i,j} n(\omega_{j},T) \bigg(\frac{\gamma_{ij}^\m{F*}}{2}\[\hat \sigma_j^\dagger \hat \rho(t), \hat \sigma_i\]+\frac{\gamma_{ji}^\m{F}}{2}\[\hat \sigma_j^\dagger, \hat \rho(t) \hat \sigma_i \]\bigg). \label{eq:dissipatorficek}
\end{align}
This equation is obtained by integrating Eq. \eqref{L} after performing the limit $\tau_R/\tau_c\to 0$ and $\tau_A/\tau_c\to\infty$ as illustrated in  Fig.\,\ref{Fig:1}\,(a). The crucial point is in the damping coefficient, $$\gamma_{ij}^\m{F}=\frac{4}{3} \frac{\vec d_i^* \cdot \vec d_j}{\hbar c^3} \omega_j^3\,,$$ which is not symmetric under exchange of the transition indices $i$ and $j$, unless they are equal. In the literature the Lindblad form is usually restored by assuming that these terms are relevant only when $\omega_i\simeq\omega_j$ and then replacing $\omega_j$ by $\sqrt{\omega_i\omega_j}$ both in $\gamma_{ij}^\m{F}$ and in the photon numbers \cite{ficek2005}, as well as replacing the corresponding terms in $\hat H_S$. Even though plausible, this procedure is arbitrary and does not allow one to estimate the error made in performing this step. In the next section, we show that these problems can be consistently solved by suitably performing the coarse-graining within the range of validity of the Born-Markov approximation. 

\subsection{Modified coarse-graining procedure}

In order to determine the coefficients of Eq.~\eqref{eq:dissipator112} and cast them in a compact form we first observe that the integration regions of Eqs.\,\eqref{H} and \eqref{L} are symmetric under the exchange of the variables $\tau_1$ and $\tau_2$. We preserve this symmetry by making the change of variables $\tau_1'=\tau_1-\tau_2$ and $\tau_2'=\tau_1+\tau_2-2t$ in the double integral of Eqs. \eqref{H}-\eqref{L}, and then by extending the integration area to the intervals $-\tau_c\le \tau_1'\le \tau_c$, $0<\tau_2'<2\tau_c$, as shown in Fig.\,\ref{Fig:1} (b). In the new integration region, the time difference $\tau_1'$ enters in physical quantities which decay with the bath correlation time $\tau_R$, whereas the contribution of the $\tau_2'$ integral can be associated with $\tau_A$.
The error performed in this operation is of the order of $\tau_R/\tau_c$ and thus within the range of validity of the derivation.

Using the modified integration region in Eq.~\eqref{eq:dissipator112} we obtain
\begin{align}
\mathcal{L}_D\tilde \rho(t)= \sum_{i,j}\vec d_i^*\cdot \vec d_j\Bigg(&\mathcal{G}\(\omega_{ij}\) \Theta_{ij}(\tau\_c) \Big(\[\tilde \sigma_j(t) \tilde \rho(t), \tilde \sigma_i^\dagger(t)\]+\[\tilde \sigma_j(t) , \tilde \rho(t) \tilde \sigma_i^\dagger(t)\]\Big) \\
+&\mathcal{G}(-\omega_{ij}) \Theta_{ij}^*(\tau\_c) \Big(\[\tilde \sigma_j^\dagger(t) \tilde \rho(t), \tilde \sigma_i(t)\]+\[\tilde \sigma_j^\dagger(t) , \tilde \rho(t) \tilde \sigma_i(t)\]\Big)\Bigg) \,,\nonumber
\end{align}
\end{widetext}
with \begin{equation}
\omega_{ij}:=(\omega_i+\omega_j)/2
\end{equation}
 and the Fourier transform of the correlation function:
 \begin{equation}
 \mathcal{G}(\omega):=\frac{1}{2}\int_{-\infty}^{\infty}\m d \tau \, C(\tau)\e^{\i \omega\tau},
 \end{equation} where the integration limits have been sent to infinity using that $\tau_c\gg\tau_R$. Moreover, we have introduced the function 
\begin{equation} 
 \Theta_{ij}(\tau\_c):=\m{exp}\big(\i(\omega_i-\omega_j)\tau\_c/2\big)\m{sinc}\big((\omega_i-\omega_j)\tau\_c/2\big).
 \end{equation}
 Specifically,
\begin{subequations}
\begin{align}
\mathcal{G}\(\omega_{ij}\) &=\frac{2}{3}\frac{1}{\hbar c^3}[1+n(\omega_{ij},T)] \omega_{ij}^3, \label{eq:Gcoefficientplus}\\
\mathcal{G}\(-\omega_{ij}\)&=\frac{2}{3}\frac{1}{\hbar c^3}n(\omega_{ij},T) \omega_{ij}^3.
\end{align}
\label{eq:Gfunctionevaluated}\end{subequations}
In order to eliminate the fast oscillations in $\Theta_{ij}$, which are an artifact of the choice of the integration over the step $\tau_c$, we integrate this function over a distribution of coarse graining times using a gaussian distribution $f(\tau\_c,\tau\_c'):=\mathcal{N} \exp(-\tau\_c'^2/\tau\_c^2)$ with a width of $\tau\_c$ and where $\mathcal{N}$ is defined so that $\int_0^{\infty} \m d \tau\_c' \,f(\tau\_c,\tau\_c')=1$. This integration smoothens the integration step and is in the spirit of the statistical meaning of the coarse graining procedure. We have checked various weighting functions and verified the convergence. After this step, we replace the function $\Theta_{ij}(\tau \_c)$ with 
\begin{equation}
\Theta_{ij}(\tau_c)\rightarrow\mathcal{F}\_c(\omega_i-\omega_j):=\int_0^{\infty}\m d \tau\_c'\, \Theta_{ij}(\tau\_c')f(\tau\_c,\tau\_c').
\end{equation}

\subsection{Cross-damping and cross-shift terms}
When moving back to the Schr\"odinger picture this procedure leads to a dissipator of the form as in Eq.~\eqref{eq:dissipatorficek}, however with the replacement
\begin{equation}
\gamma_{ij}^F\to\gamma_{ij}=\frac{4}{3}\frac{\vec d_i^*\cdot \vec d_j}{\hbar c^3}{\mathcal F}_c(\omega_i-\omega_j)\,\omega_{ij}^3\,,
\label{eq:coarsecrossdampingterms}
\end{equation}
and $$n(\omega_j,T)\rightarrow n(\omega_{ij},T)\,.$$
We remark that the fact that the frequencies here appear in the symmetric form of their arithmetic average is the result of the new integration procedure. Equation~\eqref{eq:coarsecrossdampingterms} agrees with the result of Ref.~\cite{ficek2005} in zeroth order in the parameter $(\omega_i-\omega_j)/(2\omega_i)$, while in first order its form is reminiscent to the one derived in Ref. \cite{Macovei}. For $\tau_R\ll\tau_c\ll\tau_A$, using Eq. \eqref{eq:coarsecrossdampingterms} in Eq. \eqref{eq:dissipatorficek} results in a dissipator with Lindblad form whose predictions can be compared with the one of the master equation, where the cross-damping terms are neglected. Using this procedure, moreover, the other terms due to cross-interference can be cast in terms of a self-Hamiltonian, which reads
\begin{equation}
\hat H_{\m{S}}=-\hbar\sum_{i,j}\Big[\(\Delta_{ij}^{-}+\Delta_{ij}^{T}\)\hat \sigma_i^\dagger\hat \sigma_j +\(\Delta_{ij}^{+} - \Delta_{ij}^{T}\)^* \hat \sigma_i \hat \sigma_j^\dagger\Big] \,,
\label{eq:hlamb}
\end{equation}
with $\Delta_{ij}^{\pm}$ the coarse-grained vacuum cross shift terms and $\Delta_{ij}^{T}:=\Delta_{ij}^{T-}-\Delta_{ij}^{T+}$ the coarse-grained temperature-dependent cross shift terms. Their evaluation requires a careful diagrammatic resummation which includes the high-energy contributions \cite{Karschenboim}. Nevertheless, their structure is already visible in the form one obtains in lowest order in the relativistic correction:
\begin{align}
&\Delta_{ij}^{\pm}:=\frac{2}{3}\frac{\vec d_i^*\cdot \vec d_j}{\pi \hbar c^3}{\mathcal F}\_c(\omega_i-\omega_j)\mathcal{P}\int_0^{\omega\_{cut}}\m d \omega\, \omega^3 \frac{1}{\omega\pm\omega_{ij}}\,,
\label{eq:coarsegrainedshiftterms}\\
&\Delta_{ij}^{T\pm}:=\frac{2}{3}\frac{\vec d_i^* \cdot \vec d_j}{\pi\hbar c^3}{\mathcal F}\_c(\omega_i-\omega_j)\mathcal{P}\int_0^{\omega\_{cut}}\m d \omega\, \omega^3 \frac{n(\omega,T)}{\omega\pm\omega_{ij}}\,.
\label{eq:coarsegrainedshifttermsthermal}
\end{align}
with $\mathcal{P}$  the Cauchy principal value of the integral and $\omega\_{cut}$ the cutoff frequency \cite{AgarwalNonSec,Karschenboim}. 
Their order of magnitude can be estimated by using existing data since
	\begin{align}
	\Delta_{ij}^{\pm} &\approx\frac{1}{2} \vec d_i^* \cdot \vec d_j{\mathcal F}\_c(\omega_i-\omega_j)\left( \frac{1}{|\vec d_i|^2} \Delta_{ii}^{\pm}+ \frac{1}{|\vec d_j|^2} \Delta_{jj}^{\pm}\right)\, ,
\label{eq:deltapm}
	\end{align}
and analogously for $\Delta_{ij}^{T\pm} $.

\section{Shift of resonance lines in Hydrogen Spectroscopy}
\label{Sec:3}

\begin{figure}
\begin{center}
\includegraphics{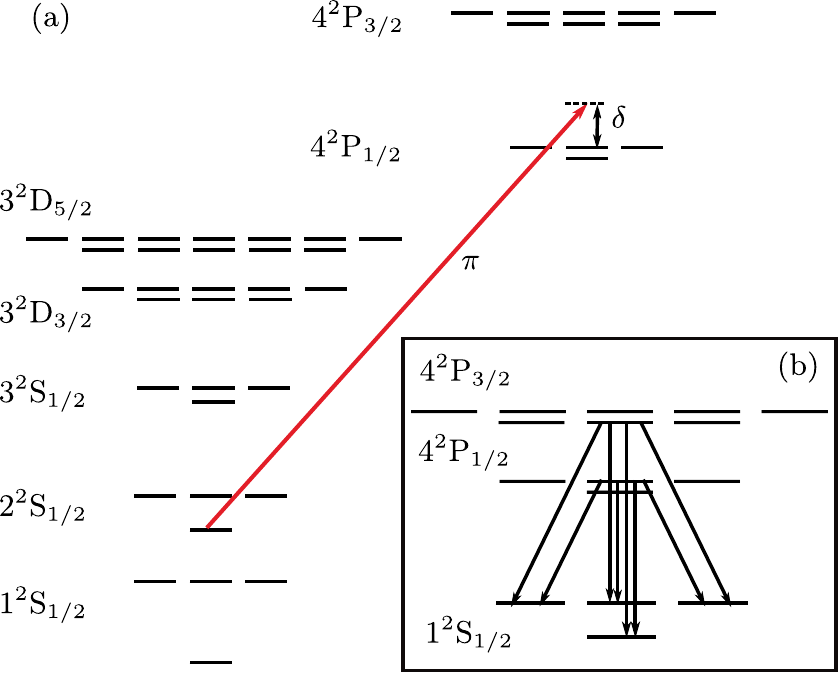}
\end{center}
\caption{(color online) (a) Relevant states for the spectroscopy of the 2S-4P transition in Hydrogen. The hyperfine structure is displayed in the standard notation where states with different $F$ ($M\_F$) quantum numbers are vertically (horizontally) displaced. The atom is assumed to be homogeneously broadened, initially prepared in the $|2^2 \mathrm{S}_{1/2}$, $F=0$, $M\_F=0\rangle$ state, and probed by a laser that is detuned by $\delta$ from the transition $|2^2 \mathrm{S}_{1/2}$, $F=0$, $M\_F=0 \rangle\to |4^2 \mathrm{P}_{1/2}$, $F=1$, $M\_F=0\rangle$. The laser is linearly polarized along $z$ and propagates along $x$. The decay channels to the 1S, 2S, 3S and 3D manifolds are taken into account and the steady state signal is obtained after a time $t=500\,\gamma\_{tot}^{-1}$, with $\gamma\_{tot}$ the total linewidth. The laser Rabi frequencies are $\sim 10^{-3}\gamma\_{tot}$. (b) Processes that lead to a vanishing line pulling when detecting over the $4\pi$ angle with no polarization filter.}
\label{fig:vanishingprocess}
\end{figure}

We test the predictions of this master equation for the spectroscopy of the $2\m{S}-4\m{P}$ transition in atomic Hydrogen in the setup of Ref. \cite{beyer2013precision}.  
The relevant states are displayed in Fig.~\ref{fig:vanishingprocess}.  The atoms are initially prepared in the state $2^2 \m{S}_{1/2},\, F=0,\, M\_F=0$. A laser, described by a classical field, probes the $2^2\m{S}-4^2\m{P}$ transition and is detuned by $\delta$ from resonance with the transition $|2^2 \mathrm{S}_{1/2}, \,F=0, \,M\_F=0\rangle\to |4^2 \mathrm{P}_{1/2}, \,F=1, \,M\_F=0\rangle$. 

For Hydrogen-like atoms, it can be shown that in the long-wavelength approximation the cross-shift terms obtained for $i\neq j$ in Hamiltonian \eqref{eq:hlamb} can be neglected~\cite{Cardimona1982steady}, as we show below. The cross-damping terms, however, lead to a distortion of the line shapes, which could induce a shift when extracting the line positions by approximating the spectra with a sum of Lorentzians as typically done in the experiment. 
A measure for this shift could be performed with the \emph{line pulling}, which we here define as 
\begin{equation}
\label{Eq:LP}
\Delta\_L:=\big[ \delta^{\m{c}}_{n} -\delta^\m{nc}_{n}\big]/(2\pi)\,,
\end{equation}
for the transition $2^2\m{S}_{\frac{1}{2}}-4^2\m{P}_{\frac{n}{2}}$ ($n=1,3$) \cite{beyer2013precision}, which is the change in the peak positions $\delta_n^\m{c}$ for the case in which the cross shift and damping terms have been incorporated and the positions $\delta_n^\m{nc}$ for when they have been set to zero. In the Appendix we show that this definition corresponds to the one of Eq.\,(12) in Ref.~\cite{jentschura2002nonresonant}.

\begin{figure}
\begin{center}
\includegraphics{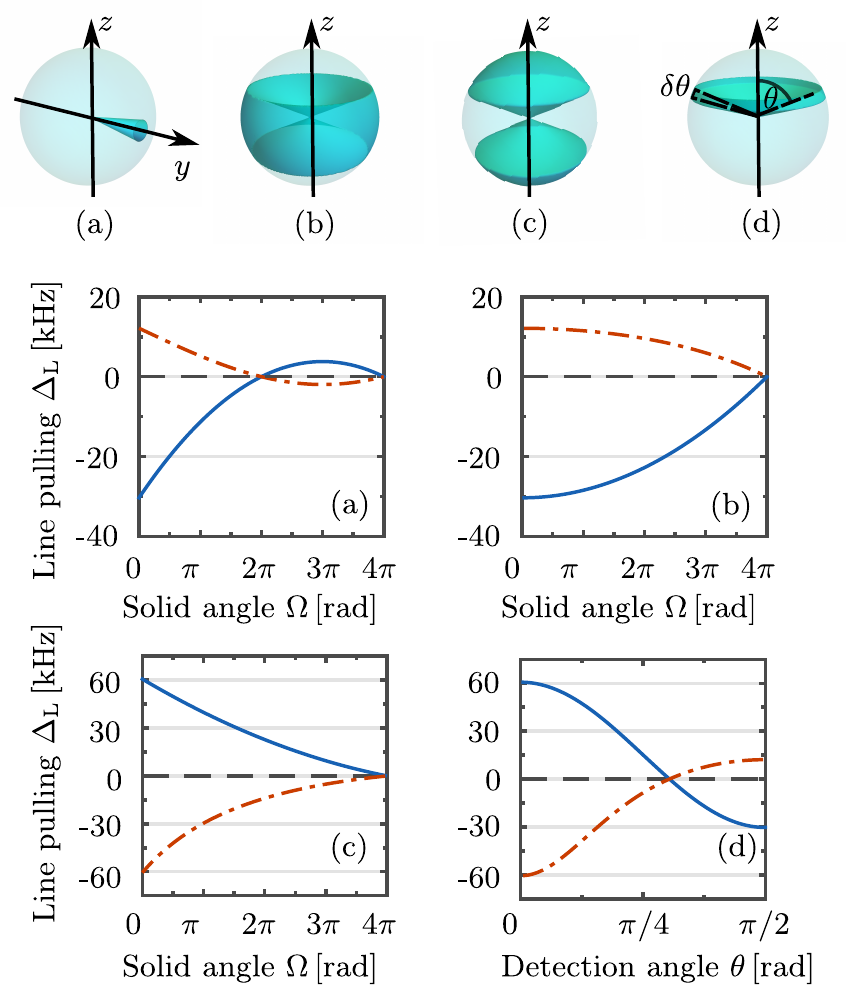}
\end{center}
\caption{(color online) Lower panels: Line pulling in the 2S-4P hyperfine transition in Hydrogen driven by a nearly resonant laser which is linearly polarized along $z$ and propagates along the $x$-axis. The different curves correspond to four different detection regions (a), (b), (c) and (d). The curves (a)-(c) are plotted as a function of the detection solid angle $\Omega$ whereas (d) is plotted as a function of the azimuthal detection angle $\theta$.The line pulling of the $4^2\m{P}_{1/2}$ ($4^2\m{P}_{3/2}$) resonance is displayed by the solid blue (dotted red) line. Upper panels: The detection regions are (a) a conic region around the $y$-axis; (b) a double cone with symmetry axis $z$, and (c) the corresponding inverted double cone (analogous to the one of Ref. \cite{beyer2013precision}). For these cases $\theta$ denotes the opening angle, while in (d) photons are detected in a stripe of width $\delta \theta$ at an azimuthal angle $\theta$.  }
\label{fig:3}
\end{figure}

\subsection{Cross-shift terms for Hydrogen-like atom}

We now show that the cross-shift terms in the self Hamiltonian, Eq. \eqref{eq:hlamb}, cancel for atomic systems that can be treated in the long-wavelength approximation if there are two interacting manifolds of states. In this case, the $\Delta_{ij}^{-}$ terms describe both a shift of and a coupling between states of the upper manifold $E=\{\ket{e}\}$. Furthermore, we only get a contribution from the operator $\hat \sigma_i^\dagger \hat \sigma_j=\ket{e}\bra{g}g'\rangle \bra{e'}$  if both transitions  share the same ground state $\ket{g}$. It is useful to split the self Hamiltonian as $$\hat H\_{S}=\hat H\_{S}^{-}+\hat H\_{S}^{+}\,,$$ where $\hat H_{\m{S}}^\pm$ includes the $\Delta_{ij}^{\pm}$ terms. We then write 
$$\hat H\_{S}^{-}=-\hbar \sum_{e,e'} \delta_{ee'} \ket{e}\bra{e'}\,,$$ where $\delta_{ee'}=\sum_{g}(\Delta_{gege'}^{-}+\Delta_{gege'}^{T})$ only contains transition operators within the upper manifold. All terms $\delta_{ee'}$ contain the function $\mathcal{F}\_c(\omega_{ge}-\omega_{ge'})$ and are proportional to terms  $\mathfrak{D}_{ee'}$\begin{equation}
\mathfrak{D}_{ee'}:=\sum_{M_g}\vec d_{ge}^* \cdot \vec d_{ge'}\,,
\label{eq:Dfrak}
\end{equation}
where the sum is over the magnetic quantum number of the ground state multiplet. We now show that $\mathfrak{D}_{ee'}$ and thus also $\delta_{ee'}$ vanishes when the states $\ket{e}$ and $\ket{e'}$, that share the same principal quantum number $n$, are different.

By evaluating the transition dipole moments using the Wigner-Eckart Theorem we obtain a formula for $\mathfrak{D}_{ee'}$ where the dependence on the magnetic quantum number $M_g$ is included in the product of two Wigner $3j$ symbols
\begin{equation}
\mathfrak{D}_{ee'}\propto \sum_{M_g,q} \begin{pmatrix} J_g & 1 & J_e \\ -M_g & q & M_e\end{pmatrix} \begin{pmatrix} J_g & 1 & J_{e'} \\ -M_g & q & M_{e'}\end{pmatrix}
\end{equation}
with $\ket{g}=\ket{n_g J_g M_g}$, $\ket{e}=\ket{n_e J_e M_e}$ and $\ket{e'}=\ket{n_{e'} J_{e'} M_{e'}}$.
Due to the orthogonality relation of the Wigner $3j$ symbols  there is only a nonvanishing contribution provided that $J_e=J_{e'}$ and $M_e=M_{e'}$ as $
\mathfrak{D}_{ee'}\propto \delta_{J_e J_{e'}}\delta_{M_e M_{e'}}$.
 If both states have the same principal quantum number, this already means that $\ket{e}=\ket{e'}$. 
We have thus proven that $\mathfrak{D}_{ee'}=0$ for $n_{e}=n_{e'}$ and $\ket{e}\neq \ket{e'}$. The product of two dipole moment vectors that connect a common ground state to two different excited states vanishes in the sum over all $M_g$ if the principal quantum number is equal for both $\ket{e}$ and $\ket{e'}$.
It follows that $\delta_{ee'}=0$ for $n_{e}=n_{e'}$ and $\ket{e}\neq \ket{e'}$.

Using an analogous procedure, the same result is obtained for the coupling between different ground states. 

\subsection{Photon count rate}

\begin{figure}
\centering
\includegraphics{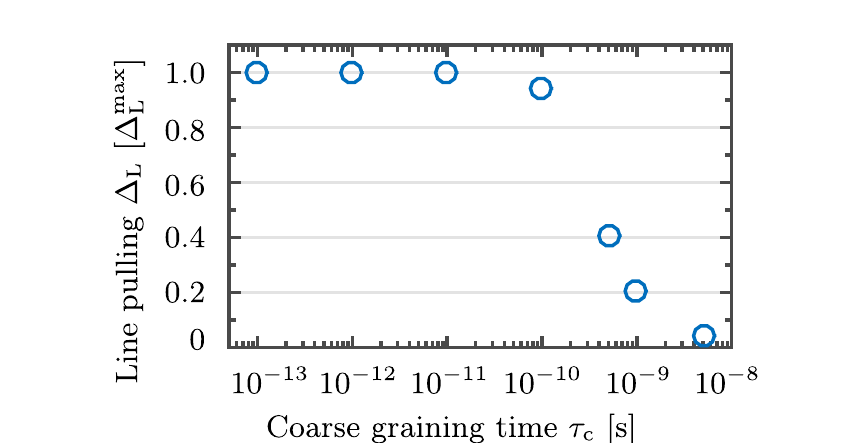}
\caption{(color online) Line pulling for the detection scheme of Fig.~2(c) and $\Omega=0$ in units of the maximum line pulling $\Delta\_L^{\m{max}}$ (indicated by the arrow in Fig.~\ref{fig:vanishingprocess}(c)) and as a function of the coarse-graining time $\tau\_c$. The line pulling remains constant in the interval $[10^{-13} ,10^{-11}]\,\m{s}$ and decreases when $\tau\_c$ approaches the inverse linewidth of the transition.}
\label{fig:4}
\end{figure}

We determine the line pulling, Eq. \eqref{Eq:LP}, from the photon count rate $S^{}(\delta)$, namely, the rate of photons emitted by an atom driven by a probe laser, as a function of the laser frequency, here given by its detuning $\delta$ from the transition $|2^2 \mathrm{S}_{1/2}$, $F=0$, $M\_F=0 \rangle\to |4^2 \mathrm{P}_{1/2}$, $F=1$, $M\_F=0\rangle$.  

In this section, we derive the photon count rate for specific detection regions. For this purpose, we first consider the expectation value of the photon number operator for a EMF mode $\lambda$ for $t'\ge t$,
$$N_\lambda(t'):=\expv{\hat a_\lambda^\dagger \hat a_\lambda}_{t'}\,,$$  where the expectation value is taken over the density matrix $\hat \chi(t')$ at time $t'$. The total photon count $N(t')$ is given by the sum of the photon counts of all modes $N(t'):=\sum_\lambda N_\lambda(t')$.  

We find an expression for $N\_{\lambda}(t')$ by following the same procedure as for the master equation. This allows us to obtain the photon count rate $$S_\lambda(t)=(N_\lambda(t+\tau\_c)-N_\lambda(t))/\tau\_c\,.$$
The total photon count rate $S^{\Omega}(t)$, namely, the rate at which photons are emitted into the solid angle $\Omega$, then depends explicitly on the cross damping terms: 
\begin{equation}
S^\Omega(t)=\sum_{i,j} \gamma_{ij}^\Omega {\rm Tr}\{\hat\sigma_j \hat \rho(t) \hat\sigma_i^\dagger \}\,,
\label{eq:photoncountratefinal}
\end{equation}
where 
\begin{equation}
 \gamma^{\Omega}_{ij}:=\frac{4}{3}\frac{\vec d_i^{\,\dagger} \overleftrightarrow{D}_\Omega \vec d_j}{\hbar c^3}  \mathcal{F}\_c(\omega_i-\omega_j)\, \omega_{ij}^3\,,
 \label{eq:gammadetection}
 \end{equation} 
and $ \overleftrightarrow{D}_\Omega=3/(8\pi)\int_\Omega (\vec e_\theta \vec e_\theta^\m{T}+\vec e_\phi \vec e_\phi^\m{T})$ is the detection matrix where the spherical coordinate vectors $\{\vec e_\theta, \vec e_\phi\}$, orthogonal to the wave vector $\vec k$, appear with the same weight, implying that we assume no polarization filter. For the case of a $4\pi$ detection angle, then $\overleftrightarrow{D}_\Omega$ reduces to  $\mathds{1}_3$ and $\gamma^{\Omega}_{ij}=\gamma_{ij}$.
 
If photons are detected over a $4\pi$ solid angle, the contribution of the cross-damping terms vanishes identically due to destructive interference between the decay channels with $\sigma^{\pm}$ and $\pi$ polarization. 
This can be understood considering that only this result can be consistent with the rotational symmetry of the Hydrogen atom. We prove it by considering the two manifolds of ground and excited states, and writing the photon count rate as $S=\sum_{ee'}G_{ee'} \rho_{ee'}$ where $G_{ee'}:=\sum_{g}\gamma^{}_{gege'}$. The terms $G_{ee'}$ are proportional to $\mathfrak{D}_{ee'}$ and to the function $\mathcal{F}\_c(\omega_i-\omega_j)$. Using the same argumentation as in the previous section, it follows that $G_{ee'}=0$ for $n_{e}=n_{e'}$ and $\ket{e}\neq \ket{e'}$. Furthermore, for states with different principal quantum numbers, the function $\mathcal{F}\_c
(\omega_i-\omega_j)$ vanishes because the difference in transition frequencies is large compared to $\tau\_c^{-1}$. Thus, for photo-detection over the full  $4\pi$ solid angle,  the cross damping terms have no measurable effect on the signal. We remark that this is true provided that there is no polarization filter.

The shift due to the cross-damping term can be different from zero in presence of a polarization filter, or  for a finite detection angle $\Omega$. We focus on the latter case for determining the line pulling observed when a beam of Hydrogen atoms is illuminated by a probe laser which drives quasi-resonantly the 2S-4P transition. Figure~\ref{fig:3} displays the line pulling as a function of the detection angle of four different detection setups. Specifically, we consider the detection solid angle $\Omega$ in subplots (a)-(c) and the detection azimuthal angle in (d). In general, the line pulling is a non-linear function of the detection angle and vanishes for specific angles. The maximum it can reach is of the order of $60\,\m{kHz}$, and this is majorly due to the contribution of the decay channel to the 1S state. The (c) detection scheme is analogous to the one implemented in Ref. \cite{beyer2013precision}. We discuss in particular the (d) detection scheme, since it shows non-trivial points at which the line pulling vanishes. Here, the photons are detected at a stripe defined by an arbitrary polar angle $\phi$ and an azimuthal angle $\theta$ with uncertainty $\delta\theta$. For a negligible width of the stripe $\delta \theta\ll 1$ one obtains lines on the unit sphere. Because of the rotational symmetry, the line pulling of the complete line then equals the line pulling of arbitrary spots on the line. For $\theta\to0,\pi$ and $\delta \theta\to 0$, namely, detection around the $z$ pole, we obtain $\Delta_L\simeq  60\, \m{kHz}$ for the resonance to $4^2\m{P}_{1/2}$ and $\Delta_L\simeq -60\, \m{kHz}$ for the $4^2\m{P}_{3/2}$ resonance. Moreover $\Delta_L$ vanishes for the angles $\theta=\m{tan}^{-1}(\sqrt{2})$ and $\theta=\pi-\m{tan}^{-1}(\sqrt{2})$ (note that $\m{tan}^{-1}(\sqrt{2})=\sin^{-1}(\sqrt{2/3})$). Most importantly, the range of the line pulling caused by the cross-damping interference includes shifts of the order or larger than $10$\,kHz, that can lead to a $4\%$ deviation of the corresponding value for the estimated r.m.s.~proton radius \cite{beyer2013precision}, and is of the order of the discrepancy between the values extracted from atomic and muonic Hydrogen spectroscopy \cite{pohl2010size,beyer2013precision,beyer2015}.

\section{Choice of the coarse-graining time}
\label{Sec:Time}

We finally analyze the dependence of the master equation's coefficients on the coarse-graining time $\tau_c$, which is the only free parameter of this theory \cite{lidar2001}. We perform an optimization following the procedure in Ref. \cite{majenz2013}. Fig.~\ref{fig:4} displays the maximum line pulling as a function of $\tau\_c$: it is constant for $\tau\_c$ in the interval $[10^{-13}\,\m s ,10^{-11}\,\m s]$, which corresponds to the range of validity of the coarse-graining. Large deviations are found for $\tau_c>10^{-10}$s, being this value too close to the atomic relaxation time: For steps of this order or larger, in fact, the coarse-graining averages out the cross-damping terms.

\section{Conclusions}
\label{Sec:Conclusions}

In conclusion, we have theoretically derived a master equation using the coarse-graining procedure, which systematically includes the terms due to cross interference in the emission into the modes of the EMF and preserves positivity, without the need of {\it ad hoc} assumptions. We applied our predictions to high-precision spectroscopy on the 2S-4P transition in Hydrogen, and showed that these cross-interference terms will have to be accurately taken into account in experiments aimed at testing the validity of quantum electrodynamics \cite{Breit}.  

Our equations further predict that dynamics due to cross interference could be better observed for atoms with no hyperfine structure and/or coupled to light in confined geometries \cite{Rauschenbeutel}. In this case also the dynamics due to cross-shifts could become visible. Extension of this treatment to other systems, where analogous interference effects can arise \cite{Antezza}, is straigthforward as long as the Markov approximation is valid. 

\begin{acknowledgments}
We thank D. Yost, and especially T. Udem and T. W. H\"ansch, and their team for motivating this work, for stimulating discussions, helpful comments, and for the critical reading of this manuscript. A.B.~acknowledges the support by the German National Academic Foundation. G.M. acknowledges illuminating discussions with P. Lambropoulos.  This work was supported by the German Research Foundation (DFG) and by the National Science Foundation under Grant No.~NSF PHY11-25915 through the Kavli Institute for Theoretical Physics at Santa Barbara (California). 
\end{acknowledgments}

\appendix
\section{Definition of the Line Pulling}
In the following, we compare the definition of the line pulling in Ref.~\cite{jentschura2002nonresonant}, Eq.~(12) to our definition given by
\begin{align}
\Delta\_L:=\big[ \delta^{\m{c}}_{n} -\delta^\m{nc}_{n}\big]/(2\pi).
\end{align}
 In the main text we approximate the photon count rate in the 2S-4P transition in Hydrogen using a sum of two Lorentzians of the form
\begin{align}
S(x)&\approx  \frac{a_1}{\pi} \frac{b_1/2}{(x-x_1)^2+(b_1/2)^2}\notag\\
&+ \frac{a_2}{\pi} \frac{b_2/2}{(x-\omega_0-x_2)^2+(b_2/2)^2},
\label{eq:sapprox}
\end{align}
where $x_1$ and $x_2$ are the approximated positions of the resonance maxima in Hz ($\delta_n:=2\pi x_n$), $b_1$ and $b_2$ are the homogeneous linewidths, $a_1$ and $a_2$ are the areas of the resonance curves and $\omega_0$ is the level splitting between the $4\m{P}_{1/2}$ and $4\m{P}_{3/2}$ states. The line pulling $\Delta\_L$ of the two resonance peaks can then be defined as
\begin{subequations}
\begin{align}
\Delta\_L(2^2 \m{S}_{1/2}-4^2 \m{P}_{1/2})&:=x_1^{\m{c}}-x_1^{\m{nc}}, \\
\Delta\_L(2^2 \m{S}_{1/2}-4^2 \m{P}_{3/2})&:=x_2^{\m{c}}-x_2^{\m{nc}},
\end{align}
\label{eq:fit1}\end{subequations}where $x_n^{\m{c}}$ are the line positions obtained from fitting spectra where the cross damping terms have been taken into account and $x_n^{ \m{nc}}$ are the spectra where all cross damping terms have been set to zero in the master equation in the main text. The results for the line pulling are displayed in Fig.~3.

We now first introduce the definition of the line pulling as found in Ref.~\cite{jentschura2002nonresonant} and then compare the line pulling presented in Fig.~3 to the values obtained using Eq.~(12) in \cite{jentschura2002nonresonant}. In Ref.~\cite{jentschura2002nonresonant} the individual resonance peaks of the photon count rate are approximated by the function
\begin{equation}
\frac{C}{x^2+\Gamma_r^2/4}+ax+\frac{b x}{x^2+\Gamma_r^2/4}=\frac{C}{[x-\Delta(x)]^2+\Gamma_r^2/4}
\end{equation}
where $\Gamma_r$ is the linewidth and $a$, $b$ and $C$ are fit parameters. $\Delta(x)$ can be approximated by
\begin{equation}
\Delta(x)=\frac{a}{2C}(x^2+\Gamma_r^2/4)^2+\frac{b}{2C}(x^2+\Gamma_r^2/4).
\label{eq:deltaj}
\end{equation}
Applying the fitting function of Eq.~\eqref{eq:deltaj} to the respective peaks in the spectrum where the cross damping terms have been taken into account, the parameters can be extracted. Ref.~\cite{jentschura2002nonresonant} gives two possible definitions of the line pulling, which we recall:

Taking 'the shift of the resonance curve at the half-maximum value as the experimentally observable measure of the apparent shift of the line center' \cite{jentschura2002nonresonant} one obtains (this is Eq.~(12) in \cite{jentschura2002nonresonant})
\begin{align}
\Delta\(\frac{\Gamma_r}{2}\)=\frac{a\Gamma_r^4}{8C}+\frac{b\Gamma_r^2}{4C},
\label{eq:j1}
\end{align}  
while taking 'the shift of the maximum of resonance' \cite{jentschura2002nonresonant} yields (this is Eq.~(13) in \cite{jentschura2002nonresonant})
\begin{align}
\Delta(0)=\frac{a\Gamma_r^4}{32 C} +\frac{b \Gamma_r^2}{8 C}.
\label{eq:j2}
\end{align}
As the terms containing the parameter $a$ in these two equations are negligible for the system under consideration, the line pulling as defined in Eq.~\eqref{eq:j1} is approximately twice as large as in \eqref{eq:j2}. 

We now apply both the definition in Eq.~\eqref{eq:j1} and our own definition in Eq.~\eqref{eq:fit1} to the same spectra that we extract from our master equation. In particular we investigate the $\theta=\pi/2$ point in Fig.~4 (d)  and compare the resulting line pullings. 

For the $4^2 \m{P}_{1/2}$ resonance we obtain
\begin{enumerate}
\item $\Delta\_L= -30 326.1\, \m{Hz}$ using our definition,
\item $\Delta\_L=-30 547.9\, \m{Hz}$ using Eq.~(12) in \cite{jentschura2002nonresonant} (respectively Eq.~\eqref{eq:j1} in this appendix).
\end{enumerate}
Moreover for the $4^2P_{3/2}$ resonance we obtain
\begin{enumerate}
\item $\Delta\_L= 12 139.5\, \m{Hz}$ using our definition,
\item $\Delta\_L=12175.9\, \m{Hz}$ using Eq.~(12) in \cite{jentschura2002nonresonant} (respectively Eq.~\eqref{eq:j1} in this appendix).
\end{enumerate}
In both cases the resulting relative deviation in the obtained value for the line pulling is smaller than 1\,\% and thus the definition in the main text and the definition using Eq.~(12) in \cite{jentschura2002nonresonant} can be considered equivalent.


\begin{thebibliography}{10}

\bibitem{Landau}
L. Landau, Zeitschrift f\"ur Physik {\bf 45},  430  (1927).

\bibitem{Milonni}
P. Milonni, Physics Reports {\bf 25},  1   (1976).

\bibitem{breuer2010theory}
H.-P. Breuer and F. Petruccione, {\em The Theory of Open Quantum Systems}
  (Oxford University Press, Oxford, 2010).

\bibitem{agarwal2012quantum}
G.~S. Agarwal, {\em Quantum Optics} (Cambridge University Press, Cambridge,
  2013).

\bibitem{Allen1975}
L. Allen and J.~H. Eberly, {\em Optical Resonance and Two-Level Atoms} (Dover
  Publications, New York, 1988).

\bibitem{pohl2010size}
R. Pohl {\it et~al.}, Nature {\bf 466},  213  (2010).

\bibitem{precisioncritique}
S.~G. Karshenboim, Phys. Rev. A {\bf 91},  012515  (2015).

\bibitem{Horbatsch2010}
M. Horbatsch and E.~A. Hessels, Phys. Rev. A {\bf 82},  052519  (2010).

\bibitem{Cardimona1983}
D.~A. Cardimona and C.~R. Stroud, Phys. Rev. A {\bf 27},  2456  (1983).

\bibitem{ficek2005}
Z. Ficek and S. Swain, {\em Quantum Interference and Coherence: Theory and
  Experiments} (Springer, New York, 2005).

\bibitem{Cook1984}
R.~J. Cook, Phys. Rev. A {\bf 29},  1583  (1984).

\bibitem{Cohen-Tannoudji1998}
C. Cohen-Tannoudji, J. Dupont-Roc, and G. Grynberg, {\em Atom-Photon
  Interactions: Basic Processes and Applications} (Wiley-VCH, Weinheim, 2004).

\bibitem{Carmichael}
H. J. Carmichael, {\it An Open Systems Approach to Quantum Optics} (Springer-Verlag, Berlin, 1993).

\bibitem{Cardimona1982steady}
D.~A. Cardimona, M.~G. Raymer, and C.~R. Stroud~Jr, J. Phys. B {\bf 15},  55
  (1982).

\bibitem{Swain}
P. Zhou and S. Swain, Phys. Rev. Lett. {\bf 77},  3995  (1996).

\bibitem{Scully1996}
S.-Y. Zhu and M.~O. Scully, Phys. Rev. Lett. {\bf 76},  388  (1996).

\bibitem{Keitel}
C.~H. Keitel, Phys. Rev. Lett. {\bf 83},  1307  (1999).

\bibitem{Berman1998}
P.~R. Berman, Phys. Rev. A {\bf 58},  4886  (1998).

\bibitem{jentschura2002nonresonant}
U.~D. Jentschura and P.~J. Mohr, Can. J. Phys. {\bf 80},  633  (2002).

\bibitem{brown2013quantum}
R.~C. Brown, S. Wu, J.~V. Porto, C.~J. Sansonetti, C.~E. Simien, S.~M. Brewer, J.~N. Tan, and J.~D. Gillaspy, Phys. Rev. A {\bf 87},  032504  (2013).

\bibitem{YostInterference2014}
D.~C. Yost, A. Matveev, E. Peters, A. Beyer, T.~W. H\"ansch, and Th. Udem, Phys. Rev. A {\bf 90},  012512  (2014).

\bibitem{lidar1999}
D. Bacon, D.~A. Lidar, and K.~B. Whaley, Phys. Rev. A {\bf 60},  1944  (1999).

\bibitem{lidar2001}
D.~A. Lidar, Z. Bihary, and K.~B. Whaley, Chem. Phys. {\bf 268},  35  (2001).

\bibitem{majenz2013}
C. Majenz, T. Albash, H.-P. Breuer, and D.~A. Lidar, Phys. Rev. A {\bf 88}, 012103  (2013).


\bibitem{schaller2008preservation}
G. Schaller and T. Brandes, Phys. Rev. A {\bf 78},  022106  (2008).

\bibitem{nonrotatingwave}
W.~J. Munro and C.~W. Gardiner, Phys. Rev. A {\bf 53},  2633  (1996).

\bibitem{AgarwalNonSec}
G.~S. Agarwal, Phys. Rev. A {\bf 7},  1195  (1973).


\bibitem{Macovei}
M. Macovei and C. H. Keitel, Phys. Rev. Lett. {\bf 91}, 123601 (2003).

\bibitem{Karschenboim}
S. G. Karschenboim, Phys. Rep. {\bf 422}, 1 (2005).

\bibitem{beyer2013precision}
A. Beyer, J. Alnis, K. Khabarova, A. Matveev, C.~G. Parthey, D.~C. Yost, R. Pohl, Th. Udem, T.~W. H\"ansch, and N. Kolachevsky, Annalen der Physik {\bf 525},  671  (2013).

\bibitem{beyer2015}
A. Beyer, L. Maisenbacher, K. Khabarova, A. Matveev, R. Pohl, Th. Udem, T.~W. H\"ansch, and N. Kolachevsky, Phys. Scr. {\bf 2015}, 014030 (2015).

\bibitem{Breit}
G. Breit, Rev. Mod. Phys. {\bf 5},  91  (1933).

\bibitem{Rauschenbeutel}
F. Le~Kien and A. Rauschenbeutel, Phys. Rev. A {\bf 90},  023805  (2014).

\bibitem{Antezza}
B. Leggio, R. Messina, and M. Antezza, EPL (Europhysics Letters) {\bf 110},
  40002  (2015).


\end{thebibliography}
\end{document}